\documentclass[12pt,aps,eqsecnum,nofootinbib,floatfix]{revtex4}
\usepackage{bbding}
\usepackage{pifont}
\usepackage{subfigure}
\usepackage{epsfig}
\usepackage{array}
\usepackage{amsmath}
\usepackage{amsfonts}
\usepackage{amssymb}
\usepackage{color}
\usepackage{graphicx}
\usepackage{mathrsfs}
\usepackage{multirow}
\usepackage{subfigure}
\usepackage[colorlinks,citecolor=blue,linkcolor=blue,urlcolor=blue,hyperindex,dvipdfm]{hyperref}

\def\be{\begin{equation}}
\def\ee{\end{equation}}
\def\bea{\begin{eqnarray}}
\def\eea{\end{eqnarray}}

\newcommand{\omits}[1]{}

\begin{document}

\title{Phase transition and entropy force between two horizons in
(n+2)-dimensional de Sitter space}
\author{Yang Zhang$^{a,b}$, Wen-qi Wang$^{a}$, Yu-bo Ma$^{a,b}$, and Jun Wang$^{c}$\footnote{\href{mailto:wjun@ynu.edu.cn}{E-mail:
wjun@ynu.edu.cn}}}
\affiliation{{\footnotesize $^a$Department of Physics, Shanxi Datong University, Datong
037009, China}\\
{\footnotesize $^b$Institute of Theoretical Physics, Shanxi Datong
University, Datong 037009, China}\\
{\footnotesize $^c$School of Physics
and Astronomy, Yunnan University, Kunming 650091, P. R. China}}

\vspace{-1.5cm}
\begin{abstract}
In this paper, the effect of the space-time dimension on effective
thermodynamic quantities in (n+2)-dimensional Reissoner-Nordstrom-de Sitter
space has been studied. Based on derived effective thermodynamic quantities,
conditions for the phase transition are obtained. The result shows that the
accelerating cosmic expansion can be attained by the entropy force arisen
from the interaction between horizons of black holes and our universe, which
provides a possible way to explain the physical mechanism for the
accelerating cosmic expansion.
\end{abstract}

\maketitle




\section{Introduction}

It is well known that the cosmic accelerated expansion indicates that our universe is a asymptotical de Sitter one. Moreover, due to the success of AdS / CFT, it prompts us to search for the similar dual relationships in de Sitter space. Therefore, the research of de Sitter space is not only of interest to
the theory itself, but also the need of the reality.

In de Sitter space, the radiation temperature on the horizon of black holes
and the universe is generally not the same. Therefore, the stability of the
thermodynamic equilibrium can not be protected in it, which makes troubles
to corresponding researches. In recent years, study on thermodynamic
properties of de Sitter space is getting more and more attention \cite%
{1,2,3,4,5,6,7,8,9,10,11,12}. In the inflationary period, our universe seem
to be a quasi de Sitter space, in which the cosmological constant is
introduced as the vacuum energy, which is a candidate for dark energy. If
the cosmological constant corresponds to dark energy, our universe will goes
into a new phase in de Sitter space. In order to construct the entire
evolutionary history of our universe, and understand the intrinsic reason
for the cosmic accelerated expansion, both the classic and quantum nature of
de Sitter space should be studied.

For a multi-horizon de Sitter space, although different horizons have
different temperatures, thermodynamic quantities on horizons of black holes
and the universe are functions depended on variables of mass, electric
charge, cosmological constant and so on. Form this point of view,
thermodynamic quantities on horizons are not individual. Based on this fact,
effective thermodynamic quantities can be introduced. Considering the
correlation between horizons of black holes and the universe, we have
studied the phase transition and the critical phenomenon in RN-dS black
holes with four-dimension and high-dimension by using effective
thermodynamic quantities, respectively. Moreover, the entropy for the
interaction between horizons of black holes and the universe is also
obtained \cite{13,14,15,16,17}. When we consider the cosmological constant
as a thermodynamic state parameter with the thermodynamic pressure, the
result shows that de Sitter space not only has a critical behavior similar
to the van der Waals system \cite{17,18}, but also take second-order phase
transition similar to AdS black hole \cite{19,20,21,22,23,24,+1,+2,+3,+4,+5}. However,
first-order phase transition similar to AdS black hole is not existed. In
this work, we investigate the issue of the phase transition in a
high-dimensional de Sitter space, and analyze the effect of the dimension on
the phase transition and the entropy produced by two interactive horizons.

Nine years ago, Verlinde \cite{25} proposed to link gravity with an entropic
force. The ensuing conjecture was proved recently \cite{26,27}, in a purely
classical environment and then extended to a quantal bosonic system in Ref.
\cite{26}. In 1998, the result of the observational data from the type Ia
supernovae (SNe Ia) \cite{40,41} indicates that our universe presently
experiences an accelerating expansion, which contrasts to the one given in
general relativity (GR) by Albert Einstein. In order to explain this
observational phenomenon, a variety of proposal have been proposed. The
theory of ``early dark energy'' proposed by Adam Riess \cite{29,30} is one
of them, where dark energy \cite{42,43} as an exotic component with large
negative pressure seems to be the cause of this observational phenomenon.
According to the observations, dark energy occupies about $73\%$ in cosmic
components. Therefore, one believe that the present accelerating expansion
of our universe should be caused by dark energy. Then a lot of dark energy
models have been proposed. However, up to now, the nature of dark energy is
not clear.

Based on the entropy caused by the interaction between horizons of black
holes and the universe, the relationship between the entropy force and the
position ratio of the two horizons is obtained. When the position ratio of
the black hole horizon to the universe horizon is greater (less) than a
certain value, the entropy force between the two horizons is repulsive
(attractive), which indicates that the expansion of the universe horizon is
accelerating (decelerating). While when it equal to the certain value, the
entropy force is absent, and then the expansion of the universe horizon is
uniform. According to this, we suppose that the entropy force between the
two horizons can be seen as a candidate to cause the cosmic accelerated
expansion.

This paper is organized as follows. According to Refs. \cite{16,17,18}, a
briefly review for the effective thermodynamic quantities, the conditions
for the phase transition and the effect of the dimension on the phase
transition in $(n+2)-$dimensional Reissoner-Nordstrom-de Sitter (DRNdS)
space is given in the next section. In section 3, the entropy force of the
interaction between horizons of black holes and the universe is derived, and
then the effect of the dimension on it is explored. Moreover, the
relationship between the entropy force and the position ratio of the two
horizons is obtained. Conclusions and discussions are given in the last
section. The units$G=\hbar =k_{B}=c=1$ are used throughout this work.

\section{Effective thermodynamic quantities}

The metric of $(n+2)-$dimensional DRNdS space is \cite{35}:
\begin{equation}
d{s^2} = - f(r)d{t^2} + {f^{ - 1}}(r)d{r^2} + {r^2}d\Omega _n^2  \label{2.1}
\end{equation}%
where the metric function is
\begin{equation}
f(r) = 1 - \frac{{{\omega _n}M}}{{{r^{n - 1}}}} + \frac{{n\omega _n^2{Q^2}}}{%
{8(n - 1){r^{2n - 2}}}} - \frac{{r^2}}{{l^2}}, {\omega _n} = \frac{{16\pi G}%
}{{nVol({S^n})}}.
\end{equation}%
Here$G$is the gravitational constant in $n+2-$dimensional space,$l$ is the
curvature radius of dS space,$Vol({S^n})$ denotes the volume of a unit $n-$%
sphere $d\Omega _n^2$,$M$ is an integration constant and $Q$ is the
electric/magnetic charge of Maxwell field.

\begin{figure}[tbp]
\centering
{\includegraphics[width=0.4\columnwidth,height=1.2in]{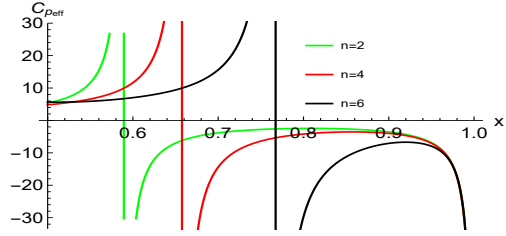}}
\par
\caption{(color online).${C_{{P_{eff}}}}-x$ diagram for $Q=0.01$,${r_{c}}=1$
and $n=2;4;6$,respectively. }
\end{figure}
\begin{figure}[tbp]
\centering
{\includegraphics[width=0.4\columnwidth,height=1.2in]{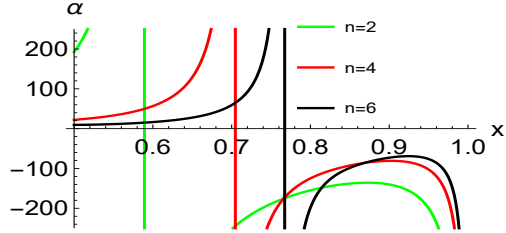}}
\par
\caption{(color online).$\protect\alpha -x$ diagram for $Q=0.01$,${r_{c}}=1$
and $n=2;4;6$,respectively. }
\end{figure}
\begin{figure}[tbp]
\centering
{\includegraphics[width=0.4\columnwidth,height=1.2in]{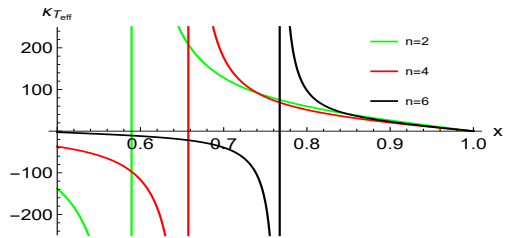}}
\par
\caption{(color online).${\protect\kappa _{{T_{eff}}}}-x$ diagram for $%
Q=0.01 $,${r_{c}}=1$ and $n=2;4;6$,respectively. }
\end{figure}

In $n+2-$dimensional DRNdS space, positions of the black hole horizon ${r_{+}%
}$ and the universe horizon ${r_{c}}$ can be determined when $f({r_{+,c}})=0$%
. Moreover, thermodynamic quantities on these two horizons satisfy the first
law of thermodynamics, respectively \cite{3,5,35}. However, thermodynamic
systems denoted by the two horizons are not independent, since thermodynamic
quantities on them are functions depended on variables of mass $M$, electric
charge $Q$ and cosmological constant ${l^{2}}$ satisfy the first law of
thermodynamics. When parameters of state of $n+2-$dimensional DRNdS space
satisfy the first law of thermodynamics, the entropy is \cite{16,17,18}
\begin{equation}
S=\frac{{Vol({S^{n}})}}{{4G}}r_{c}^{n}(1+{x^{n}}+{f_{n}}(x))={S_{c,+}}+{%
S_{AB}},
\end{equation}%
where $x={r_{+}}/{r_{c}}$,${S_{c,+}}=\frac{{Vol({S^{n}})}}{{4G}}r_{c}^{n}(1+{%
x^{n}})$ and ${S_{AB}}=\frac{{Vol({S^{n}})}}{{4G}}r_{c}^{n}{f_{n}}(x)$ are
entropies with and without the interaction between the two horizons,
respectively, and
\begin{equation}
{f_{n}}(x)=\frac{{3n+2}}{{2n+1}}{(1-{x^{n+1}})^{n/(n+1)}}-\frac{{(n+1)(1+{%
x^{2n+1}})+(2n+1)(1-2{x^{n+1}}-{x^{2n+1}})}}{{(2n+1)(1-{x^{n+1}})}}.
\end{equation}%
The volume of $n+2-$dimensionalDRNdS space is\cite{3,7,13}
\begin{equation}
V={V_{c}}-{V_{+}}=\frac{{Vol({S^{n}})}}{{(n+1)}}r_{c}^{n+1}(1-{x^{n+1}}).
\end{equation}%
When parameters of state of $n+2-$ dimensional DRNdS space satisfy the first
law of thermodynamics, the effective temperature is \cite{16,17,18}
\begin{eqnarray}
{T_{eff}} &=&(1-{x^{n+1}})\frac{{{{(\partial M/\partial x)}_{{r_{c}}}}(1-{%
x^{n+1}})+{r_{c}}{x^{n}}{{(\partial M/\partial {r_{c}})}_{x}}}}{{Vol({S^{n}}%
)r_{c}^{n}{x^{n-1}}(1+{x^{n+2}})}}  \notag \\
&=&\frac{{B(x)}}{{Vol({S^{n}}){r_{c}}{x^{2n-1}}{\omega _{n}}(1+{x^{n+2}})}},
\end{eqnarray}%
where
\begin{eqnarray}
B(x) &=&{x^{n}}[(n-1){x^{n-2}}-(n+1){x^{n}}+2{x^{2n-1}}+(n-1){x^{2n-1}}(1-{%
x^{2}})]  \notag \\
&&-\frac{{n\omega _{n}^{2}{Q^{2}}[(n-1){x^{n+1}}(1-{x^{2n}})-2n{x^{n+1}}%
+(n-1)+(n+1){x^{2n}}]}}{{8(n-1)r_{c}^{2n-2}}}  \notag \\
&=&{x^{n}}[(n-1){x^{n-2}}-(n+1){x^{n}}+2{x^{2n-1}}+(n-1){x^{2n-1}}(1-{x^{2}}%
)]  \notag \\
&&-\frac{{2\phi _{c}^{2}(n-1)[(n-1){x^{n+1}}(1-{x^{2n}})-2n{x^{n+1}}%
+(n-1)+(n+1){x^{2n}}]}}{n},
\end{eqnarray}%
where ${\phi _{c}}=\frac{n}{{4(n-1)}}\frac{{{\omega _{n}}Q}}{{r_{c}^{n-1}}}$
is electric potential on the universe horizon. The effective pressure $%
P_{eff}$, isochoric heat capacity $C_{veff}$ and isobaric heat capacity $%
C_{P_{veff}}$ in $n+2-$dimensional DRNdS spaceare
\begin{equation}
{P_{eff}}=\frac{{D(x)}}{{{\omega _{n}}Vol({S^{n}})(1-{x^{n+1}})r_{c}^{2}{%
x^{n-1}}(1+{x^{n+2}})}},
\end{equation}%
where
\begin{eqnarray}
D(x) &=&\left[ {(n-1){x^{n-2}}-(n+1){x^{n}}+2{x^{2n-1}}-\frac{{n\omega
_{n}^{2}{Q^{2}}(2n{x^{n+1}}-(n-1)-(n+1){x^{2n}})}}{{8(n-1)r_{c}^{2n-2}{x^{n}}%
}}}\right] \times   \notag \\
&&(1+{x^{n}}+f(x)) \\
&&-\left[ {(n-1){x^{n-1}}(1-{x^{2}})-\frac{{n\omega _{n}^{2}{Q^{2}}(1-{x^{2n}%
})}}{{8r_{c}^{2n-2}{x^{n-1}}}}}\right] \left( {{x^{n-1}}+\frac{{f^{\prime
}(x)}}{n}}\right) (1-{x^{n+1}}),  \notag
\end{eqnarray}%
\begin{eqnarray}
{C_{V}} &=&{T_{eff}}{\left( {\frac{{\partial S}}{{\partial {T_{eff}}}}}%
\right) _{V}}={T_{eff}}\frac{{{{\left( {\frac{{\partial S}}{{\partial {r_{c}}%
}}}\right) }_{x}}{{\left( {\frac{{\partial V}}{{\partial x}}}\right) }_{{%
r_{c}}}}-{{\left( {\frac{{\partial S}}{{\partial x}}}\right) }_{{r_{c}}}}{{%
\left( {\frac{{\partial V}}{{\partial {r_{c}}}}}\right) }_{x}}}}{{{{\left( {%
\frac{{\partial V}}{{\partial x}}}\right) }_{{r_{c}}}}{{\left( {\frac{{%
\partial {T_{eff}}}}{{\partial {r_{c}}}}}\right) }_{x}}-{{\left( {\frac{{%
\partial V}}{{\partial {r_{c}}}}}\right) }_{x}}{{\left( {\frac{{\partial {%
T_{eff}}}}{{\partial x}}}\right) }_{{r_{c}}}}}} \\
&=&\frac{1}{{4G(1-{x^{n+1}})}}\times   \notag \\
&&\frac{{-Vol({S^{n}})r_{c}^{n}B(x)n{x^{n}}{{(1+{x^{n+2}})}^{2}}}}{{{\bar{B}%
(x){x^{n+1}}(1+{x^{n+2}})-(1-{x^{n+1}}){x(1+{x^{n+2}})B^{\prime
}(x)-B(x)[2n-1+(3n+1){x^{2n+2}}]}}}}  \notag
\end{eqnarray}%
where
\begin{eqnarray}
\bar{B}(x) &=&{x^{n}}[(n-1){x^{n-2}}-(n+1){x^{n}}+2{x^{2n-1}}+(n-1){x^{2n-1}}%
(1-{x^{2}})]  \notag \\
&&-\frac{{n\omega _{n}^{2}{Q^{2}}(2n-1)[(n-1){x^{n+1}}(1-{x^{2n}})-2n{x^{n+1}%
}+(n-1)+(n+1){x^{2n}}]}}{{8(n-1)r_{c}^{2n-2}}},  \notag \\
B^{\prime }(x) &=&\frac{{dB(x)}}{{dx}},D^{\prime }(x)=\frac{{dD(x)}}{{dx}},
\end{eqnarray}%
\begin{eqnarray}
{C_{{P_{eff}}}} &=&{T_{eff}}{\left( {\frac{{\partial S}}{{\partial {T_{eff}}}%
}}\right) _{{P_{eff}}}}={T_{eff}}\frac{{{{\left( {\frac{{\partial S}}{{%
\partial {r_{c}}}}}\right) }_{x}}{{\left( {\frac{{\partial {P_{eff}}}}{{%
\partial x}}}\right) }_{{r_{c}}}}-{{\left( {\frac{{\partial S}}{{\partial x}}%
}\right) }_{{r_{c}}}}{{\left( {\frac{{\partial {P_{eff}}}}{{\partial {r_{c}}}%
}}\right) }_{x}}}}{{{{\left( {\frac{{\partial {P_{eff}}}}{{\partial x}}}%
\right) }_{{r_{c}}}}{{\left( {\frac{{\partial {T_{eff}}}}{{\partial {r_{c}}}}%
}\right) }_{x}}-{{\left( {\frac{{\partial {P_{eff}}}}{{\partial {r_{c}}}}}%
\right) }_{x}}{{\left( {\frac{{\partial {T_{eff}}}}{{\partial x}}}\right) }_{%
{r_{c}}}}}}  \notag \\
&=&r_{c}^{n}\frac{{Vol({S^{n}})B(x)E(x)}}{{4GH(x)}},
\end{eqnarray}%
where
\begin{eqnarray}
E(x) &=&[n{x^{n-1}}+f^{\prime }(x)][\bar{D}(x)-2D(x)](1-{x^{n+1}})x(1+{%
x^{n+2}})  \notag \\
&&-n[1+{x^{n}}+f(x)]\{D^{\prime }(x)x(1-{x^{n+1}})(1+{x^{n+2}})  \notag \\
&&-D(x)[(n-1)-2n{x^{n+1}}+(2n+1){x^{n+2}}-(3n+2){x^{2n+3}}]\},  \notag \\
H(x) &=&\bar{B}(x)\{D^{\prime }(x)x(1-{x^{n+1}})(1+{x^{n+2}})-D(x)[(n-1)
\notag \\
&&-2n{x^{n+1}}+(2n+1){x^{n+2}}-(3n+2){x^{2n+3}}]\} \\
&&+(1-{x^{n+1}})[\bar{D}(x)-2D(x)]\left[ {x(1+{x^{n+2}})B^{\prime
}(x)-B(x)[2n-1+(3n+1){x^{2n+2}}]}\right] .  \notag \\
\bar{D}(x) &=&\frac{{n\omega _{n}^{2}{Q^{2}}(2n{x^{n+1}}-(n-1)-(n+1){x^{2n}})%
}}{{4r_{c}^{2n-2}{x^{n}}}}(1+{x^{n}}+f(x))  \notag \\
&&-\frac{{n(n-1)\omega _{n}^{2}{Q^{2}}(1-{x^{2n}})}}{{4r_{c}^{2n-2}{x^{n-1}}}%
}\left( {{x^{n-1}}+\frac{{f^{\prime }(x)}}{n}}\right) (1-{x^{n+1}}).  \notag
\end{eqnarray}%
The coefficient of isobaric volume expansion and isothermal compressibility
in $n+2-$ dimensional DRNdS spaceis given by
\begin{eqnarray}
\alpha  &=&\frac{1}{V}{\left( {\frac{{\partial V}}{{\partial {T_{eff}}}}}%
\right) _{{P_{eff}}}}=\frac{1}{V}\frac{{{{\left( {\frac{{\partial V}}{{%
\partial {r_{c}}}}}\right) }_{x}}{{\left( {\frac{{\partial {P_{eff}}}}{{%
\partial x}}}\right) }_{{r_{c}}}}-{{\left( {\frac{{\partial V}}{{\partial x}}%
}\right) }_{{r_{c}}}}{{\left( {\frac{{\partial {P_{eff}}}}{{\partial {r_{c}}}%
}}\right) }_{x}}}}{{{{\left( {\frac{{\partial {P_{eff}}}}{{\partial x}}}%
\right) }_{{r_{c}}}}{{\left( {\frac{{\partial {T_{eff}}}}{{\partial {r_{c}}}}%
}\right) }_{x}}-{{\left( {\frac{{\partial {P_{eff}}}}{{\partial {r_{c}}}}}%
\right) }_{x}}{{\left( {\frac{{\partial {T_{eff}}}}{{\partial x}}}\right) }_{%
{r_{c}}}}}}  \notag \\
&=&-\frac{{{\omega _{n}}(n+1)Vol({S^{n}}){x^{2n-1}}(1+{x^{n+2}})}}{{H(x)}}{%
r_{c}}\{{x^{n+1}}[\bar{D}(x)-2D(x)](1+{x^{n+2}}) \\
&&+D^{\prime }(x)x(1-{x^{n+1}})(1+{x^{n+2}})-D(x)[(n-1)-2n{x^{n+1}}+(2n+1){%
x^{n+2}}-(3n+2){x^{2n+3}}]\}.  \notag
\end{eqnarray}%
\begin{eqnarray}
{\kappa _{{T_{eff}}}} &=&-\frac{1}{V}{\left( {\frac{{\partial V}}{{\partial {%
P_{eff}}}}}\right) _{{T_{eff}}}}=\frac{1}{V}\frac{{{{\left( {\frac{{\partial
V}}{{\partial {r_{c}}}}}\right) }_{x}}{{\left( {\frac{{\partial {T_{eff}}}}{{%
\partial x}}}\right) }_{{r_{c}}}}-{{\left( {\frac{{\partial V}}{{\partial x}}%
}\right) }_{{r_{c}}}}{{\left( {\frac{{\partial {T_{eff}}}}{{\partial {r_{c}}}%
}}\right) }_{x}}}}{{{{\left( {\frac{{\partial {P_{eff}}}}{{\partial x}}}%
\right) }_{{r_{c}}}}{{\left( {\frac{{\partial {T_{eff}}}}{{\partial {r_{c}}}}%
}\right) }_{x}}-{{\left( {\frac{{\partial {P_{eff}}}}{{\partial {r_{c}}}}}%
\right) }_{x}}{{\left( {\frac{{\partial {T_{eff}}}}{{\partial x}}}\right) }_{%
{r_{c}}}}}}  \notag \\
&=&\frac{{r_{c}^{2}{\omega _{n}}(n+1)Vol({S^{n}})(1-{x^{n+1}}){x^{n-1}}(1+{%
x^{n+2}})}}{{H(x)}}\times  \\
&&\left\{ {(1-{x^{n+1}})\left[ {x(1+{x^{n+2}})B^{\prime }(x)-B(x)[2n-1+(3n+1)%
{x^{2n+2}}]}\right] -{x^{n+1}}(1+{x^{n+2}})\bar{B}(x)}\right\}   \notag
\end{eqnarray}

Numerical solutions for the isobaric heat capacity $C_{p_{eff}}$ and
coefficients of isobaric volume expansion $\alpha$ and isothermal
compressibility $\kappa_{T_{eff}}$ with the position ratio of the black hole horizon to the universe horizon $x$ have been given in Fig. 1, Fig. 2 and Fig. 3, respectively. Form the figures, it is clear that values of
$C_{p_{eff}}$, $\alpha$ and $\kappa_{T_{eff}}$ have sudden change with
the charge of the spacetime is a constant, which is similar to the Van der
Waals system. Moreover, as the dimension of the space increases, the value
of $x$ to denote the sudden change also increases. This indicates that the
point of the phase transition is closely related to the dimensions of the
space time.

%
%
%
\begin{table*}
 \begin{center}
 \caption[]{\it Critical values of the effective thermodynamic system
for different $n$}
 \begin{tabular}{|c|c|c|c|c|c|c|} \hline
 \cline{1-4}
   &  $ n=2 $   & $n=4 $   & $n=6$ \\
        \hline
           $ x_c $     & 0.5894    &0.7053    & 0.7674\\
        \hline
           $ T_{eff}^c $    & 0.0301   & 0.1127   & 0.2095\\
        \hline
           $ P_{eff}^c$   & 0.0238    & 0.0952    & 0.1825\\
        \hline
 \end{tabular}
 \end{center}
 \label{tab:1}
 \end{table*}

From Table 1, it is clear that the phase transition point is different with
different dimensions. Moreover, as the dimension increases, the critical
value of the phase transition point and the effective pressure and
temperature are all increased.

\section{Entropy force}

The entropy force of a thermodynamic system can be expressed as \cite%
{25,26,27,36,37,38,39}
\begin{equation}
F = - T\frac{{\partial S}}{{\partial r}},
\end{equation}
where $T$ is the temperature and $\gamma$ is the radius.

From Eq.(2.3), the entropy caused by the interaction between horizons of
black holes and the universe is
\begin{equation}
{S_{AB}} = \frac{{Vol({S^n})}}{{4G}}r_c^n{f_n}(x).
\end{equation}
\begin{figure}[tbp]
\centering
{\includegraphics[width=0.4\columnwidth,height=1.2in]{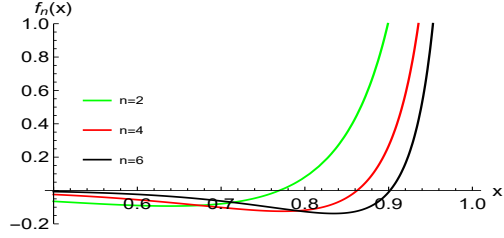}}
\par
\caption{(color online).${f_n}(x) - x$ diagram for $\frac{{Vol({S^n})}}{{4G}}%
\mathrm{\ = }1,r_c^n = 1$ and $n = 2;4;6$,respectively. }
\end{figure}

From Fig.4, it shows that as the dimension increases, the intersectional
point of the curve and the $x-$axis is moving to the right.In other words,
the value of ${x_0}$ increases with the dimension,which denotes the point
where the entropy caused by the interaction between horizons of black holes
and the universe changes between positive and negative values. The entropy
given in Eq. (2.4) does not contain explicit electric charge Q dependent $Q$
terms.

From Eq. (3.1), the entropy force of the two interactive horizons can be
given as
\begin{equation}
F=-{T_{eff}}{\left( {\frac{{\partial {S_{AB}}}}{{\partial r}}}\right) _{{%
T_{eff}}}},
\end{equation}%
where ${T_{eff}}$ is the effective temperature of the considering case and $%
r={r_{c}}-{r_{+}}={r_{c}}(1-x)$. Then it gives
\begin{eqnarray}
F(x) &=&-{T_{eff}}\frac{{{{\left( {\frac{{\partial {S_{f}}}}{{\partial {r_{c}%
}}}}\right) }_{x}}{{\left( {\frac{{\partial {T_{eff}}}}{{\partial x}}}%
\right) }_{{r_{c}}}}-{{\left( {\frac{{\partial {S_{f}}}}{{\partial x}}}%
\right) }_{{r_{c}}}}{{\left( {\frac{{\partial {T_{eff}}}}{{\partial {r_{c}}}}%
}\right) }_{x}}}}{{(1-x){{\left( {\frac{{\partial {T_{eff}}}}{{\partial x}}}%
\right) }_{{r_{c}}}}+{r_{c}}{{\left( {\frac{{\partial {T_{eff}}}}{{\partial {%
r_{c}}}}}\right) }_{x}}}} \\
&=&\frac{{-B(x)r_{c}^{n-2}}}{{4G{x^{2n-1}}{\omega _{n}}(1+{x^{n+2}})}}\times
\notag \\
&&\frac{{n{f_{n}}(x)\left[ {x(1+{x^{n+2}})B^{\prime }(x)-B(x)[2n-1+(3n+1){%
x^{2n+2}}]}\right] +x(1+{x^{n+2}})\bar{B}(x)f{_{n}^{\prime }}(x)}}{{(1-x)%
\left[ {x(1+{x^{n+2}})B^{\prime }(x)-B(x)[2n-1+(3n+1){x^{2n+2}}]}\right] +{%
x^{2}}\bar{B}(x)(1+{x^{n+2}})}}.  \notag
\end{eqnarray}%
\begin{figure}[tbp]
\centering
{\includegraphics[width=0.4\columnwidth,height=1.2in]{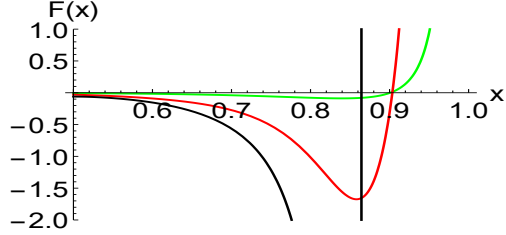}}
\par
\caption{(color online).$F(x)-x$ diagram for $Q=0.01$,${r_{c}}=1$ and $%
n=2;4;6$,respectively.}
\end{figure}

Fig.5 shows that the entropy force increases with the dimension. Moreover,
when $n=2$ and $x={x_{0}}=0.9009$, $n=4$ and $x={x_{0}}=0.9035$ , and $n=6$
and $x={x_{0}}=0.9224$ , $F({x_{0}})=0$, respectively. It indicates that the
value of ${x_{0}}$ increases with the dimension, which denotes the point
where the direction of the entropy force changes.

\begin{figure}[tbp]
\centering
{\includegraphics[width=0.4\columnwidth,height=1.2in]{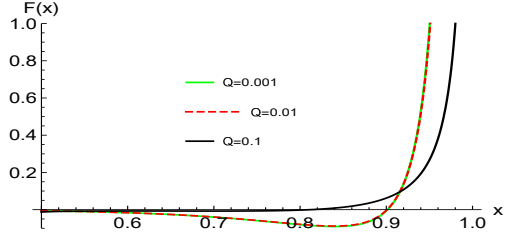}}
\par
\caption{(color online).$F(x) - x$ diagram for $n = 2$,${r_c} = 1$ and $Q =
0.001;0.01;0.1$,respectively.}
\end{figure}

Fig. 6 shows that when $Q=0.001$ and $x={x_{0}}=0.9014,$ $Q=0.01$ and $x={%
x_{0}}=0.9009$, and $Q=0.1$ and $x={x_{0}}=0.8120,F({x_{0}})=0$
respectively. It implies that as the electric charge increases, the value of
${x_{0}}$ decreases, which denotes the point where the entropy force changes
between positive and negative values.

From Fig. 5, we can obtain that when $x\rightarrow 1,$ $F(x\rightarrow
1)\rightarrow \infty $, and then according to Eq. (2.6), ${T_{eff}}%
\rightarrow 0$. This result indicates that the interaction between horizons
of black holes and the universe tends to infinity, which contrast to the
third law of thermodynamics. In order to protect the laws of thermodynamics,
the black hole horizon and the cosmological horizon can not coincide with
each other. Based on this fact, we take $1-\Delta x$ as the maximum value of
$x$, where $\Delta x$ is a minor dimensionless quantity. The value of $%
\Delta x$ can be determined by the speed of the cosmic accelerated expansion
at the position $x$.

According to the expression of the entropy force, when ${x_{0}}<x<1-\Delta
x, $ ${F}(x)>0$ which indicates that the interaction between horizons of
black holes and the universe is repulsive. Consequently,the expansion of the
cosmological horizon can be accelerated by the entropy force in the absence
of other forces. In Fig. 5, it is known that the entropy force is different
at different positions. Thus the expansion of the universe is variable
acceleration in the interval of ${x_{0}}<x<1-\Delta x$. While when $0<x<{%
x_{0}}$, ${F}(x)<0$, which indicates that the interaction between horizons
of black holes and the universe is attractive, and then the expansion of the
universe is variable deceleration in this interval.

From Fig. 5, we find that when the area enclosed by the curve $F(x)-x$ and
the $x-$axis with the interval of ${x_{0}}<x<1-\Delta x$ is larger than the
area enclosed by the same curve and the $x-$axis with the interval of $0<x<{%
x_{0}}$,the cosmic expansion is from acceleration to deceleration. It gives
an expanding universe. While when the former area is less than or equal to
the latter one, the cosmic expansion is from acceleration to deceleration.
Moreover, when these two areas are equal at the position ratio $x$ , which
belongs to the interval of $\bar{x}<x<{x_{0}},$ the universe is accelerated
shrinkage from the position ratio $\bar{x}$ to the position ratio ${x_{0}}$,
where $\bar{x}$ is determined when the area between the curve and the x-axis
with the interval of $[\bar{x},1-\Delta x]$ is zero. After the
universeshrink to the position ratio $x=1-\Delta x,$ the evolution of the
universe begins the next cycle. It gives a oscillating universe.

\section{Conclusions}

When horizons of black holes and the universe are irrelevant, thermodynamic
systems of them are independent. Since the radiational temperature on them
is different, the requirement of thermodynamic equilibrium stability can not
be meet. Therefore, the space is unstable. While when they are related, the
effective temperature ${T_{eff}}$ and pressure ${P_{eff}}$ for DRNdS space
can be obtained from Eqs.(2.6) and (2.8). According to curves ${C_{{P_{eff}}}%
}-x$, $\alpha -x$, and ${\kappa _{{T_{eff}}}}-x$, when $x={x_{c}}$, the
phase transition of DRNdS space time occurs. Since its entropy and volume are
continuous, the phase transition is the second-order one according to
Ehrenfest's classification. It is similar to the case occured in AdS black
holes \cite{19,20,21,22,23,24,44,45}. From Eq. (2.10), we find that the
isochoric heat capacity ${C_{v}}$ of DRNdS space is non-trivial, which is
similar to the system of Van der Waals, but different from AdS black
holes. In second 2, the effect of the dimension on the phase transition point
is analyzed, which lays the foundation for the further study of the
thermodynamic characteristics of the high-dimensional complex dS space.

From Fig. 5, we find that when the area enclosed by the curve $F(x)-x$ and
the $x-$axis with the interval of ${x_{0}}<x<1-\Delta x$ is larger than the
area enclosed by the same curve and the $x-$axis with the interval of $0<x<{%
x_{0}}$, the cosmic expansion is from acceleration to deceleration. It gives
an expanding universe. While when the former area is less than or equal to
the latter one, the cosmic expansion is from acceleration to deceleration.
Moreover, when these two areas are equal at the position ratio $x$, which
belongs to the interval of $\bar{x}<x<{x_{0}}$, the universe is accelerated
shrinkage from the position ratio $\bar{x}$ to the position ratio ${x_{0}}$,
where $\bar{x}$ is determined when the area between the curve and the $x-$%
axis with the interval of $[\bar{x},1-\Delta x]$ is zero. After the universe
shrink to the position ratio $x=1-\Delta x$, the evolution of the universe
begins the next cycle. It gives a oscillating universe.

Whether the universe is an expanding one or a oscillating one is determined
by the value of the minor dimensionless quantity. From Fig. 5 and Fig.6, we
find that the position, where the entropy force changes between positive and
negative values, is greatly affected by the dimension, but commonly by the
electric charge. Therefore, the effect of the dimension on the cosmic
expansion is greater than the electric charge. Moreover, since the curve $%
F(x) - x$ is continuous at the phase transition point ${x_c}$, the entropy
force can not be affected by the phase transition in the space with a given
dimension and electric charge. The amplitude and the value of the entropy
force is only determined by the position ratio $x$. According to our
research result, the entropy force between horizons of black holes and the
universe can be taken as one of the reasons for the cosmic expansion, which
provides a new approach for people to explore the physical mechanism of the
cosmic expansion.

\section*{Acknowledgements}
We thank Prof. Z. H. Zhu for useful discussions.
This work was supported by the National Natural Science Foundation of China
(Grant No. 11847123, 11847128, 11475108, 11705106, 11705107, 11605107),
Science $\&$ Technology Department of Yunnan Province - Yunnan University Joint Funding (Grant No. 2019FY003005) and Donglu Youth Teacher Plan of Yunnan University.


\begin{thebibliography}{99}
\bibitem{1} S. Mbarek, R. B. Mann, Reverse Hawking-Page Phase
Transition in de Sitter Black Holes. JHEP 02, 103 (2019).

\bibitem{2} F. Simovic, R. B. Mann, Critical Phenomena of Charged de
Sitter Black Holes in Cavities. Classical and Quantum Gravity, 36, 014002
(2019).

\bibitem{3} B. P. Dolan, D. Kastor, D. Kubiznak, R. B. Mann,
Jennie Traschen, Thermodynamic Volumes and Isoperimetric Inequalities for de
Sitter Black Holes, Phys. Rev. D 87, 104017 (2013).

\bibitem{4} S. H. Hendi, A. Dehghani, Mir Faizal, Black hole thermodynamics
in Lovelock gravity's rainbow with (A)dS asymptote, Nucl. Phys. B 914, 117
(2017).

\bibitem{5} Y. Sekiwa, Thermodynamics of de Sitter black holes: Thermal
cosmological constant. Phys. Rev. D 73, 084009 (2006).

\bibitem{6} D. Kubiznak, F. Simovic, Thermodynamics of horizons: de
Sitter black holes and reentrant phase transitions. Classical and Quantum
Gravity,33, 245001(2016).

\bibitem{7} J. McInerney, G. Satishchandran, J. Traschen,
Cosmography of KNdS Black Holes and Isentropic Phase Transitions, Classical
and Quantum Gravity,33, 105007(2016).

\bibitem{8} M. Urano and A. Tomimatsu, Mechanical First Law of Black
Hole Spacetimes with Cosmological Constant and Its Application to
Schwarzschild-de Sitter Spacetime, Classical and Quantum Gravity,26,
105010(2009).

\bibitem{9} S. Bhattacharya, A. Lahiri, Mass function and particle
creation in Schwarzschild-de Sitter spacetime, Eur. Phys. J. C 73,2673(2013).

\bibitem{10} R. G. Cai, Gauss-Bonnet black holes in AdS, Phys. Rev. D
65,084014(2002).

\bibitem{11} A. Kumar Chatterjee, K. Flathmann, H. Nandan and A. Rudra,
Analytic solutions of the geodesic equation fo rReissner-Nordstrom
Anti-de-Sitter black holes surrounded bydifferent kinds of regular and
exotic matter fields. Arxiv: 1903.11878

\bibitem{12} L. C. Zhang, R. Zhao, The critical phenomena of
Schwarzschild-de Sitter Black hole, EPL,113,10008(2016).

\bibitem{13} L. C. Zhang, M. S. Ma, H. H. Zhao, R. Zhao, Thermodynamics of
phase transition in higher dimensional Reissner-Nordstrom-de Sitter black
hole, The European Physical Journal C. 74, 3052 (2014).

\bibitem{14} H. H. Zhao, L. C. Zhang, M. S. Ma, and R. Zhao, PV criticality
of higher dimensional charged topological dilaton de Sitter black holes.
Phys. Rev. D 90, 064018 (2014).

\bibitem{15} H. F. Li, M. S. Ma, L. C. Zhang, and R. Zhao, Entropy of
Kerr-de Sitter black hole, Nuclear Physics B 920, 211-220(2017).

\bibitem{16} L. C. Zhang, R. Zhao, M. S. Ma, Entropy of
Reissner--Nordstrm--de Sitter black hole, 761,74--76 (2016).

\bibitem{17} R. Zhao and L. C. Zhang, Entropy of Higher-Dimensional Charged
de Sitter Black Holes and Phase Transition, Commun. Theor. Phys.70,
578--584(2018).

\bibitem{18} Y. B. Ma, L. C. Zhang, T. Peng, Y. Pan and S. Cao, Entropy of
the electrically charged hairy black holes, The European Physical Journal C,
78,763(2018).

\bibitem{19} S. H. Hendi, R. B. Mann, S. Panahiyan, B. Eslam Panah, van der
Waals like behaviour of topological AdS black holes in massive gravity.
Phys. Rev. D 95, 021501(R) (2017).

\bibitem{20} R. A. Hennigar, E. Tjoa, R. B. Mann, Thermodynamics of hairy
black holes in Lovelock gravity. JHEP 2017, 70(2017).

\bibitem{21} R. A. Hennigar, R. B. Mann, E. Tjoa, Superfluid Black Holes.
Phys. Rev. Lett. 118, 021301 (2017).

\bibitem{22} A. M. Frassino, D. Kubiznak, R. B. Mann, F. Simovic, Multiple
Reentrant Phase Transitions and Triple Points in Lovelock Thermodynamics.
JHEP 09, 080(2014).

\bibitem{23} D. Kubiznak, R. B. Mann, P-V criticality of charged AdS black
holes, JHEP 07, 033 (2012).

\bibitem{24} R. G. Cai, L. M. Cao, L. Li, and R. Q. Yang, \textquotedblleft
P-V criticality in the extended phase space of Gauss-Bonnet black holes in
AdS space\textquotedblright , JHEP 09, 005 (2013).


\bibitem{+1} X. X. Zeng and Y. W. Han, Holographic van der Waals Phase Transition for a Hairy Black Hole, Advances in High Energy Physics 2017,  2356174(2017).
    
\bibitem{+2} X. X. Zeng, L. F. Li, Holographic phase transition probed by non-local observables, Advances in High Energy Physics 2016,  6153435 (2016).
    
\bibitem{+3} S. He, L. F. Li, X. X. Zeng, Holographic Van der Waals-like phase transition in the Gauss-Bonnet gravity, Nuclear Physics B 915, 243(2017).
    
\bibitem{+4} X. X. Zeng, X. M. Liu, L. F. Li, Phase structure of the Born-Infeld-anti-de Sitter black holes probed by non-local observables, Eur. Phys. J. C 76, 616(2016).
    
\bibitem{+5} X. X. Zeng, H. B. Zhang, L. F. Li, Phase transition of holographic entanglement entropy in massive gravity, Physics Letters B 756, 170 (2016). 

\bibitem{25} E. P. Verlinde, On the Origin of Gravity and the Laws of
Newton, JHEP 04, 029 (2011).

\bibitem{26} A. Plastino, M. C. Rocca, G. L. Ferri, Quantum treatment of
Verlinde's entropic force conjecture, Physica A 511, 139 (2018).

\bibitem{27} C. P. Panos, Ch.C. Moustakidis, A simple link of information
entropy of quantum and classical systems with Newtonian ${r^{ - 2}}$
dependence of Verlinde's entropic force, Physica A 518, 384 (2019).

\bibitem{29} A. G. Riess, L. Macri, S. Casertano., et al., A 3\% Solution:
Determination of the Hubble Constant with the Hubble Space Telescope and
Wide Field Camera 3, The Astrophysical Journal, 730 , 119 (2011).

\bibitem{30} A. G. Riess, A. Filippenko, P. Challis., et al. Observational
Evidence from Supernovae for an Accelerating Universe and a Cosmological
Constant. The Astronomical Journal, 116, 1009 (1998).

\bibitem{35} R .G. Cai, Cardy-Verlinde formula and thermodynamics of black
holes in de Sitter spaces. Nucl. Phys. B 628, 375 (2002).

\bibitem{36} Y. F. Cai, J. Liu, H. Li, Entropic cosmology: A unified model
of inflation and late-time acceleration, Phys. Lett, B 690,213-219 (2010).

\bibitem{37} Z. Q. Zhang, Z. J. Luo, D. F. Hou, Higher derivative
corrections to the entropic force from holography, Annals of Physics 391,
47-55 (2018).

\bibitem{38} Z. Q. Zhang, Z. J. Luo, D. F. Hou, G. Chen, Entropic
destruction of heavy quarkonium from a deformed AdS model, arXiv:1701.06147

\bibitem{39} D. E. Kharzeev, Deconfinement as an entropic self-destruction:
A solution for the quarkonium suppression puzzle?, Phys. Rev. D 90, 074007
(2014).

\bibitem{40} B. P. Schmidt, N. B. Suntzeff, M. M. Phillips, et al. The
High-Z Supernova Search: Measuring Cosmic Deceleration and Global Curvature
of the Universe Using Type Ia Supernovae. Astrophys. J. 507(1)46 (1998).

\bibitem{41} S. Perlmutter, G. Aldering, G. Goldhaber, et al. Measurements
of $\Omega$ and $\Lambda$ from 42 High-Redshift Supernovae. Astrophys. J.
517(2);565 (1999).

\bibitem{42} P. J. E. peebles, B. Ratra. The cosmological constant and dark
energy. Rev. Mod. Phys, 75(2),559-606 (2003).

\bibitem{43} S. Wang, Y. Wang, M. Li. Holographic dark energy. Phys. Rept,
696,1-57 (2017).

\bibitem{44} W. Xu, C. Y. Wang, and B. Zhu, Effects of Gauss-Bonnet term on
the phase transition of a Reissner-Nordstr?m-AdS black hole in (3+1)
dimensions, Phys. Rev. D 99, 044010 (2019).

\bibitem{45} S. W. Wei, Y. X. Liu, and Y. Q. Wang, Probing the relationship
between the null geodesics and thermodynamic phase transition for rotating
Kerr-AdS black holes, Phys. Rev. D 99, 044013 (2019).
\end{thebibliography}
\end{document}